\documentclass[prb,twocolumn,superscriptaddress]{revtex4-1}

\usepackage{amsmath}
\usepackage{amssymb}
\usepackage{xspace}
\usepackage{graphicx}
\usepackage{grffile}
\usepackage{nicefrac}
\usepackage{color}

\graphicspath{{figs/}}

\begin{document}

\title{Oxygen-vacancy driven electron localization and itinerancy in rutile-based TiO$_2$}

\author{Frank Lechermann}
\affiliation{I. Institut f{\"u}r Theoretische Physik, Universit{\"a}t Hamburg, 
D-20355 Hamburg, Germany}
\affiliation{Institut f{\"u}r Keramische Hochleistungswerkstoffe, Technische Universit\"at
Hamburg-Harburg, D-21073 Hamburg, Germany}
\author{Wolfgang Heckel}
\affiliation{Institut f{\"u}r Keramische Hochleistungswerkstoffe, Technische Universit\"at
Hamburg-Harburg, D-21073 Hamburg, Germany}
\author{Oleg Kristanovski}
\affiliation{I. Institut f{\"u}r Theoretische Physik, Universit{\"a}t Hamburg, 
D-20355 Hamburg, Germany}
\author{Stefan M\"uller}
\affiliation{Institut f{\"u}r Keramische Hochleistungswerkstoffe, Technische Universit\"at
Hamburg-Harburg, D-21073 Hamburg, Germany}


\begin{abstract}
Oxygen-deficient TiO$_2$ in the rutile structure as well as the Ti$_3$O$_5$ Magn{\'e}li phase 
is investigated within the charge self-consistent combination of density functional theory 
(DFT) with dynamical mean-field theory (DMFT). It is shown that an isolated 
oxygen vacancy (V$_{\rm O}$) in titanium dioxide is not sufficient to metallize the system at 
low temperatures. In a semiconducting phase, an in-gap state is identified at 
$\varepsilon_{\rm IG}^{\hfill}\sim -0.75\,$eV\, in excellent agreement with experimental data.
Band-like impurity levels, resulting from a threefold V$_{\rm O}$-Ti coordination as well
as entangled $(t_{2g},e_g)$ states, become localized due to site-dependent electronic correlations.
Charge localization and strong orbital polarization occur in the V$_{\rm O}$-near Ti ions, 
which details can be modified by a variation of the correlated subspace.
At higher oxygen vacancy concentration, a correlated metal is stabilized in the Magn{\'e}li 
phase. A V$_{\rm O}$-defect rutile structure of identical stoichiometry shows key differences in 
the orbital-resolved character and the spectral properties. Charge disproportionation is vital 
in the oxygen-deficient compounds, but obvious metal-insulator transitions driven or sustained 
by charge order are not identified.
\end{abstract}

\maketitle

\section{Introduction}
From two motivating research directions, the investigation of oxygen-deficient 
transition-metal oxides has gained enormous renewed interest. First,
the emerging field of oxide heterostructures lead to questions concerning
the impact of oxygen vacancies on interface properties. Since especially the SrTiO$_3$ 
band insulator marks an important heterostructure building block, elucidating the role 
of such vacancies in that compound has recently attracted lots of 
attention.~\cite{luo04,pav12,mit12,she12,lec14,lec16,alt16}
Second, on the search of realizing a memristor,~\cite{yan13} TiO$_{2-\delta}$ 
remains a key material.~\cite{kwo10,hec15} Formation and migration of oxygen-vacancy 
defects are identified to regulate the resistance modulation therein.

\begin{figure}[b]
\includegraphics*[width=6cm]{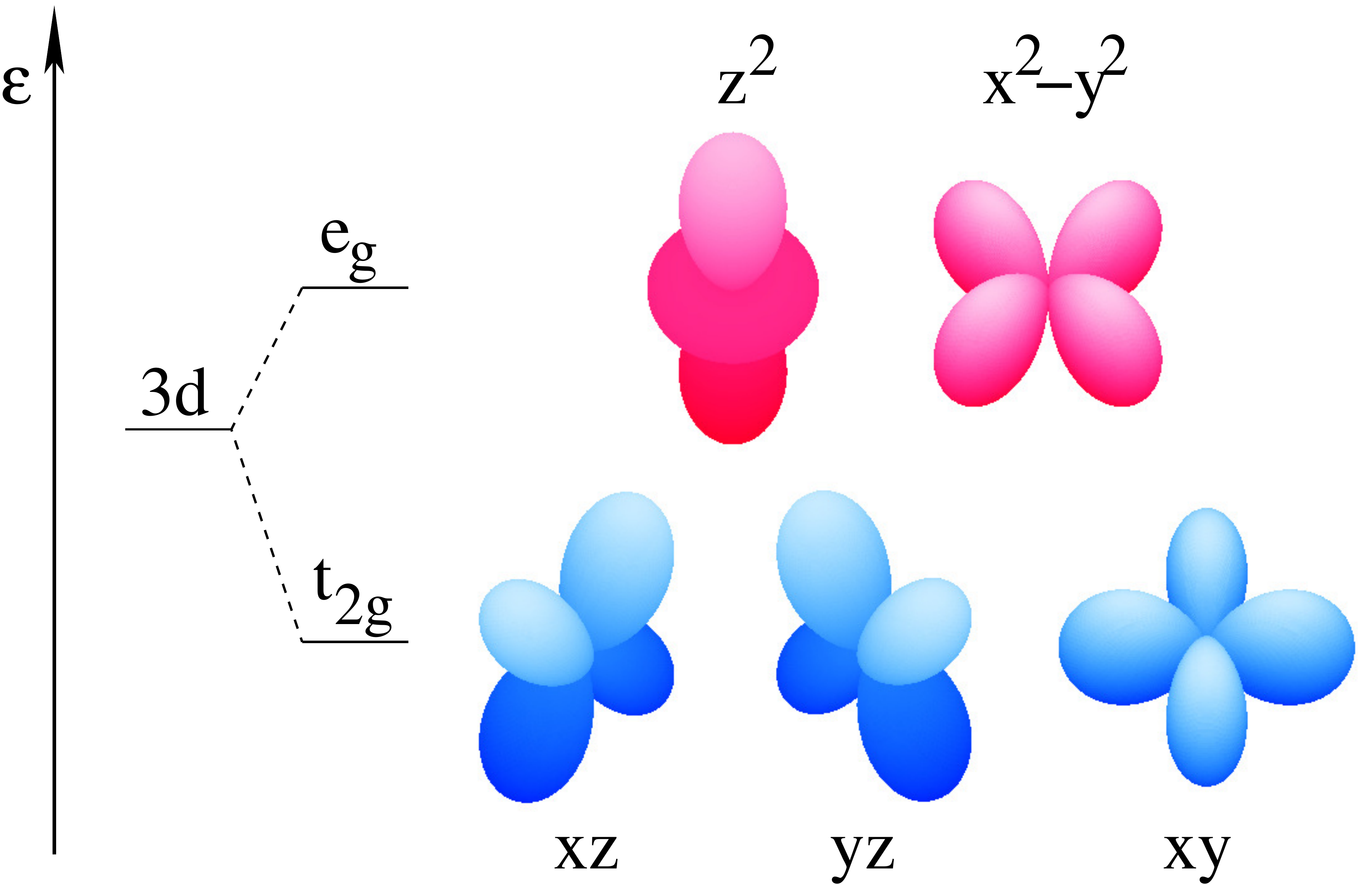}
\caption{(color online) Crystal-field splitting of a transition-metal
$3d$ shell, located in a full-cubic symmetry environment, into $t_{2g}$ and 
$e_g$ states.}
\label{fig:dsplit}
\end{figure}

Stoichiometric SrTiO$_3$ is characterized as a cubic (perovskite) Ti$^{4+}(3d^0)$ 
compound with crystal-field split $e_g$ and $t_{2g}$ states (cf. Fig.~\ref{fig:dsplit}).
The band gap is located between the dominantly O$(2p)$- and the $t_{2g}$ manifold. Due to 
the strong O$(2p)-e_g$  hybridization, the creation of an oxygen vacancy (V$_{\rm O}$) 
leads to local Ti$^{3+}(3d^1)$ sites and $e_g$-dominated in-gap states.~\cite{luo04} 
The interplay of Ti$^{4+}$- and Ti$^{3+}$-like states gives 
rise to a competition between electron localization and itinerancy, posing an 
intriguing many-body problem. Recently, that problem was approached by theory
within the combination of density functional theory (DFT) with dynamical mean-field 
theory (DMFT).~\cite{lec14,beh15,lec16} Experiments indeed suggest that V$_{\rm O}$s on the 
surface of strontium titanate as well as in the interface of LaAlO$_3$/SrTiO$_3$ may be 
relevant not only for metallicity, but also for emergent magnetic and/or 
superconducting order.

Oxygen vacancies in TiO$_2$ pose a related intricate problem, yet with a twist. Besides 
single-defect scenarios, long-range-ordered vacancy structures provided by the 
Ti$_n$O$_{2n-1}$ Magn{\'e}li phases~\cite{asb59,and63,bur72} are an additional 
point of materials reference.
The role of oxygen vacancies in titanium dioxide, with its twofold structural 
representations of rutile and anatase is a longstanding problem, and has so far been 
studied in several theoretical works based on conventional DFT,~\cite{lib08} using hybrid 
functionals,~\cite{jan10,dea12,jan13,ber15,vas16} as well as by treating static electronic correlations 
within DFT+U.~\cite{mat08,mat10,mor10,lin15,vas16} Within the latter framework, 
Mattioli {\sl et al.}~\cite{mat08} originally showed that isolated V$_{\rm O}$s in TiO$_2$ can 
introduce shallow electronic levels only in anatase, while solely deep localized 
levels are induced in rutile. The more recent oxygen-deficient rutile 
studies~\cite{mat10,jan13,lin15} suggest an intricate coexistence of shallow and deep levels.

Electronic structure investigations of various Magn{\'e}li phases, which may in fact be 
derived starting from the rutile structure,~\cite{lib08} furthermore revealed challenging 
physics, such as e.g. metal-insulator transitions and charge ordering.~\cite{leo06,pad14,tan15} 
But those assessments are so far limited by the possibilities of DFT(+U) to describe electron 
correlation.

In this work we want to provide a theoretical account of electron correlations 
in oxygen-deficient rutile-based TiO$_2$ from a DFT+DMFT perspective. This not only 
provides a relevant examination of defect-mediated electronic self-energy effects beyond 
Kohn-Sham exchange-correlation treatments for an oxide compound 
with high potential for technological applications. It also allows us to compare the 
characteristics of the induced defect states to e.g. the ones found in the SrTiO$_3$ 
perovskite. We show that energy-similar in-gap states are emerging upon creation of 
V$_{\rm O}$s, but metallicity does only occur above a corresponding concentration 
threshold.

Rutile is known to be the thermodynamically stable TiO$_2$ phase at all temperatures and 
pressures,~\cite{han11} while anatase is metastable but can kinetically be stabilised
at low temperatures. We therefore restrict the investigation on the rutile structural 
case of TiO$_2$ as well as the Ti$_3$O$_5$ Magn{\'e}li phase as a higher 
V$_{\rm O}$-concentration counterpart.

\section{Computational Approach}
The supercell defect structures of rutile-TiO$_{2-\delta}$ as well as the Magn{\'e}li
Ti$_3$O$_5$ structure are structurally relaxed~\cite{hec15} on the DFT level within the 
generalized-gradient approximation (GGA) using the PBEsol~\cite{per08} functional in 
the VASP code.~\cite{kre93,kre94,kre96-1,kre96-2}

Our present charge self-consistent DFT+DMFT framework~\cite{sav01,pou07,gri12} builds up
on the mixed-basis pseudopotential approach~\cite{lou79,mbpp_code} for the DFT part and 
the continuous-time quantum-Monte-Carlo method~\cite{rub05,wer06}, as implemented in 
the TRIQS package,~\cite{par15,set16} for the DMFT impurity problem. We utilize the
GGA in the PBE~\cite{per96} functional form within the Kohn-Sham cycle.

Locally, threefold effective Ti$(3d)$ functions define the correlated subspace, which
as a whole consists of the corresponding sum over the various Ti sites in the defect problem.
Projected-local orbitals~\cite{ama08,ani05,aic09,hau10,kar11} of $3d$ character provide the
effective functions from acting on Kohn-Sham conduction states above the O$(2p)$-dominated
band manifold. Note that the resulting effective orbitals are not of exclusive $t_{2g}$-
or $e_g$ kind, but are defined by the local three-orbital sector lowest in energy, 
respectively. Each Ti site marks an impurity problem, and the whole number
of explicitly treated impurity problems depends on the number of symmetry-inequivalent
transition-metal sites in the given supercell. A three-orbital Hubbard Hamiltonian of 
Slater-Kanamori form, if not otherwise stated parametrized by the Hubbard $U=5\,$eV and 
the Hund's exchange $J_{\rm H}=0.7\,$eV, is active on each Ti site. These values for the 
local Coulomb interactions are in line with previous studies on 
titanates.~\cite{miz95,pav04,oka06,lec15}
A double-counting correction of the fully-localized form~\cite{ani93} is utilized in 
this work. The analytical continuation of the finite-temperature Green's functions on the 
Matsubara axis $i\omega$ to real frequencies is performed via the maximum-entropy method.

Note that we study only paramagnetic states without broken spin symmetry, as well as
charge-neutral lattice configurations. Yet possible inter-site charge disproportionation 
is surely allowed.

\section{Results}
\subsection{Rutile TiO$_2$}
\begin{figure}[b]
\parbox[c]{5.5cm}{
(a)\includegraphics*[width=5cm]{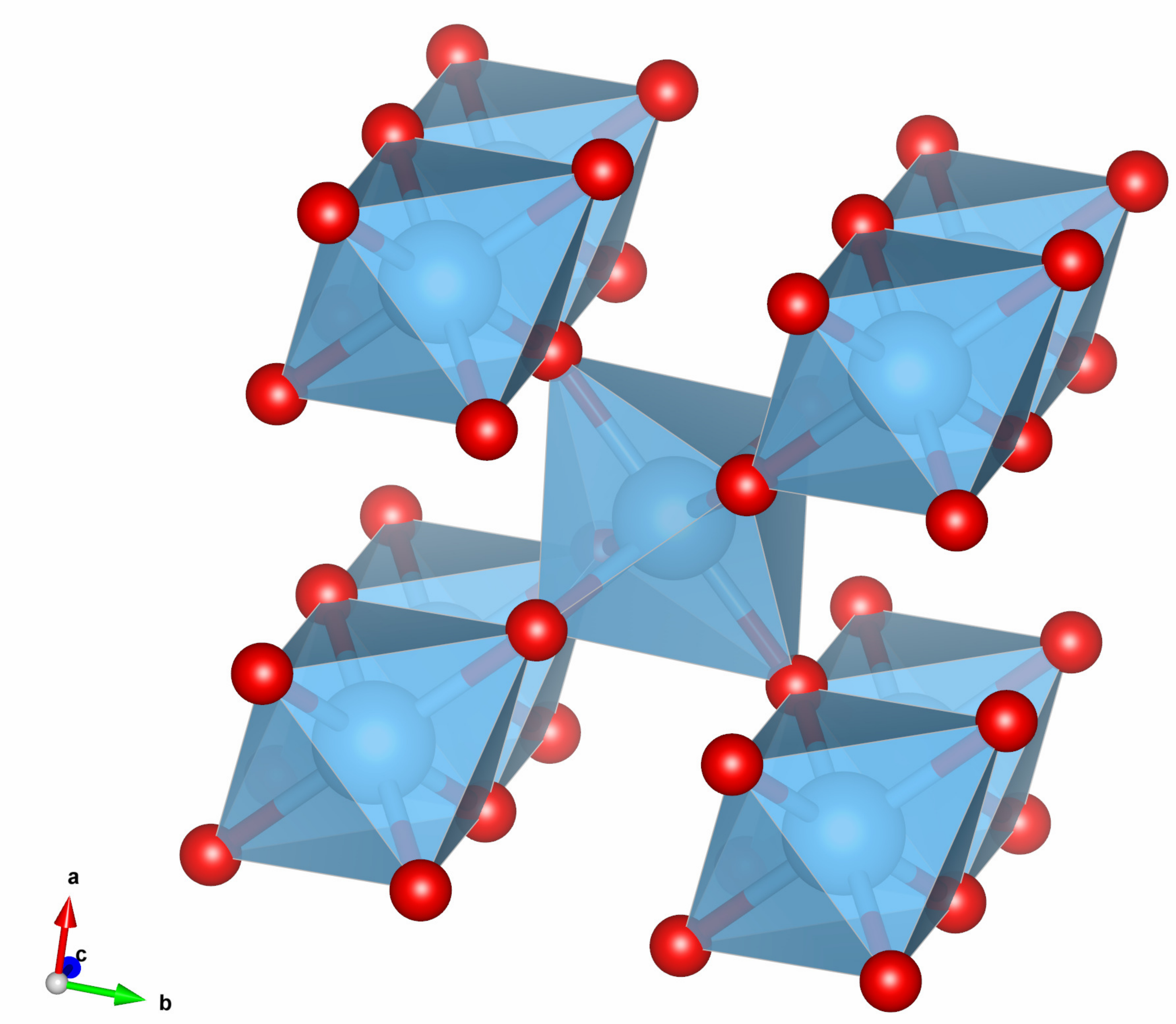}\\[0.2cm]
(b)\hspace*{-0.3cm}\includegraphics*[width=5.15cm]{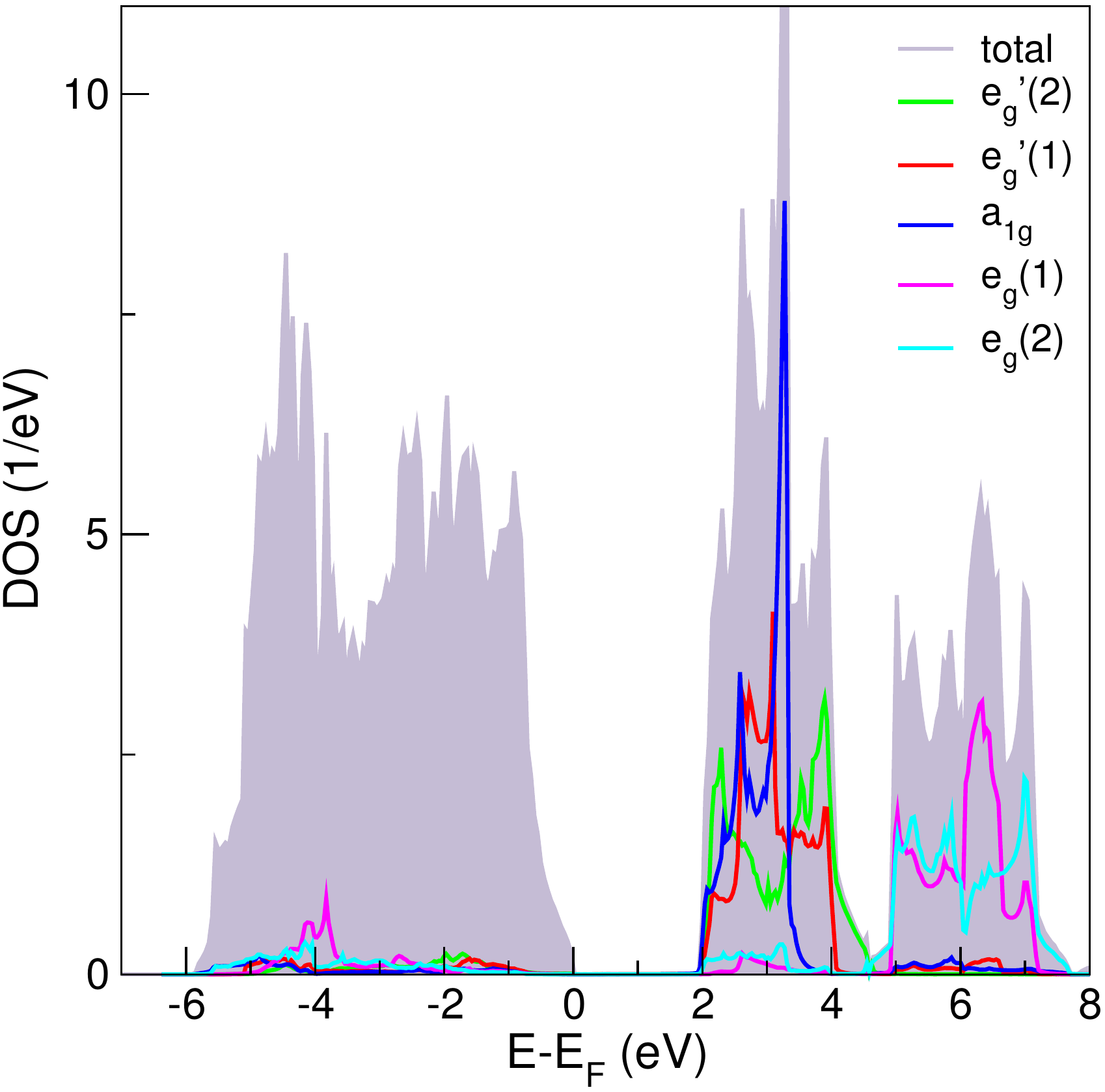}}
\parbox[c]{2.75cm}{(c)\includegraphics*[width=2.5cm]{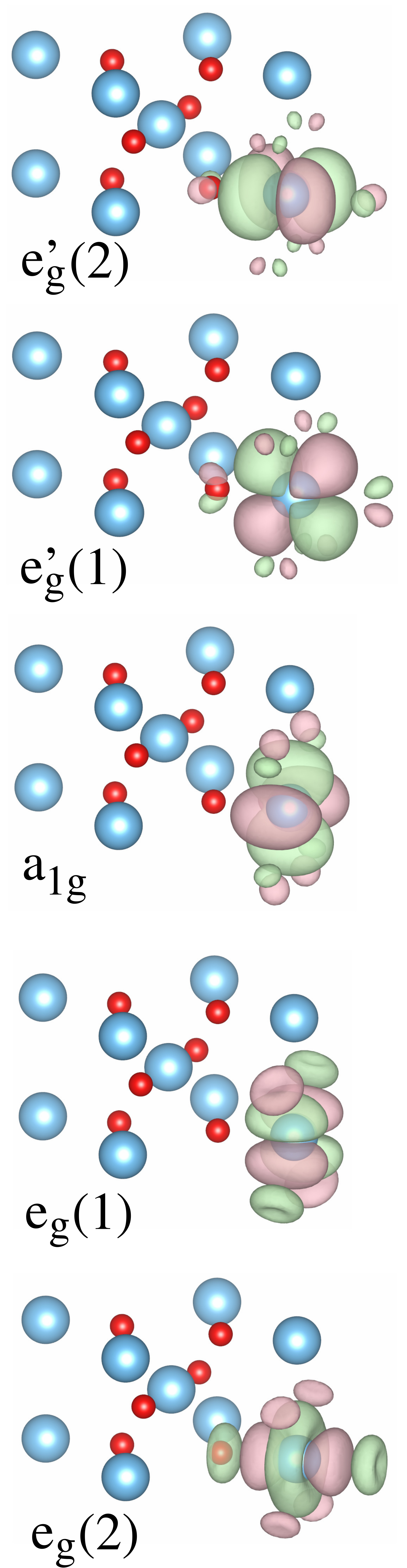}}
\caption{(color online) Characterization of rutile-TiO$_2$.
(a) Crystal structure with Ti (large blue/grey) and O (small red/dark) atoms.
(b) Total and local-orbital GGA density of states.
(c) Projected local Ti$(3d)$ orbitals.  
}\label{fig:erutile}
\end{figure}
To set the stage for the discussion of oxygen-deficient titanium dioxide, we briefly discuss
the electronic structure at stoichiometry. The rutile structure~\cite{bur87} 
(see Fig.~\ref{fig:erutile}a) has tetragonal symmetry (space group $P4_2/mnm$) with a 
ratio $c/a=0.64$ and the primitive cell comprises two TiO$_2$ formula units. 
It consistes of corner- and edge-sharing TiO$_6$ octahedra, such that each oxygen ion is 
coordinated by three neighboring titanium ions. This is in contrast to common 
perovskite-based titanates, where the TiO$_6$ octahedra are exclusively corner-sharing and O 
is twofold Ti-coordinated. Of the three rutile minimal Ti-O bond lengths, the two shorter 
ones are identical in extent. Nominally, titanium is in the Ti$^{4+}$ oxidation state with
$3d^0$ occupation.

The given compound is a band insulator with an experimental (optical) band gap of size 
$\Delta_{\rm g}\sim 3$\,eV.~\cite{cro51,cro52,pas78} Conventional DFT calculations yield a 
smaller gap $\Delta_{\rm g}\sim 2$\,eV (cf. Fig.~\ref{fig:erutile}b). Since TiO$_2$ is
a nominal $3d^0$ material, the HOMO orbitals are of dominant O$(2p)$ kind and the LUMO orbitals 
are mainly of Ti$(3d_{t_{2g}})$ character, while dominant Ti$(3d_{e_{g}})$ is located at even higher
energies above the band gap. Because of the fact that the $c$-axis and the main TiO$_6$-octahedra
axes are locally aligned trigonal, the Ti($t_{2g},e_g$) orbitals may be written as 
linear combinations of cubic harmonics from diagonalization of the orbital-density matrix 
(see Tab.~\ref{tab:rut}).
\begin{table}[t]
\begin{ruledtabular}
\begin{tabular}{l|cccccc}
orbital   & $\varepsilon^{\hfill}_{\rm CF}$ & $|z^2\rangle$ & $|xz\rangle$ & $|yz\rangle$ & $|x^2-y^2\rangle$ 
& $|xy\rangle$  \\ \hline
$|e_g'(2)\rangle$  & 2684  &  0.000 &  0.000 & 0.000 & 1.000 &  0.000 \\
$|e_g'(1)\rangle$ & 2653  &  0.000 &  0.707 & 0.707 & 0.000 &  0.000 \\
$|a_{1g}\rangle$ & 2574  & -0.827 &  0.000 & 0.000 & 0.000 & -0.562 \\
$|e_g(1)\rangle$  & 4398  &  0.000 & -0.707 & 0.707 & 0.000 &  0.000 \\
$|e_g(2)\rangle $ & 4413  &  0.562 &  0.000 & 0.000 & 0.000 & -0.827
\end{tabular}
\end{ruledtabular}
\caption{The titanium $t_{2g}=(e_g'(2),e_g'(1),a_{1g})$ and $e_g=(e_g(1),e_g(2))$ orbitals
in TiO$_2$ with their respective crystal-field level $\varepsilon^{\hfill}_{\rm CF}$ (in meV),
expressed in terms of cubic harmonics.}
\label{tab:rut}
\end{table}
Due to the local symmetry, the internal $(t_{2g},e_g)$ degneracies known from the full octahedral
group are lifted, respectively. Comparison of the crystal-field levels $\varepsilon^{\hfill}_{\rm CF}$ 
marks the $a_{1g}$ level as the lowest one, 110 meV below $e_g'(2)$.
The $t_{2g}$-based states are about $2.2$\,eV lower in energy than the $e_g$-based ones. 
Within the effective $t_{2g}$ manifold, which has a bandwdith of about $2.5$\,eV, the $e_g'(2)$ 
orbital is designated since its lobes point along the in-plane tetragonal axes 
(see Fig.~\ref{fig:erutile}c).

\subsection{Oxygen vacancy in rutile TiO$_2$}
\subsubsection{Structural details and correlated subspaces}
\begin{figure}[t]
\includegraphics*[width=6cm]{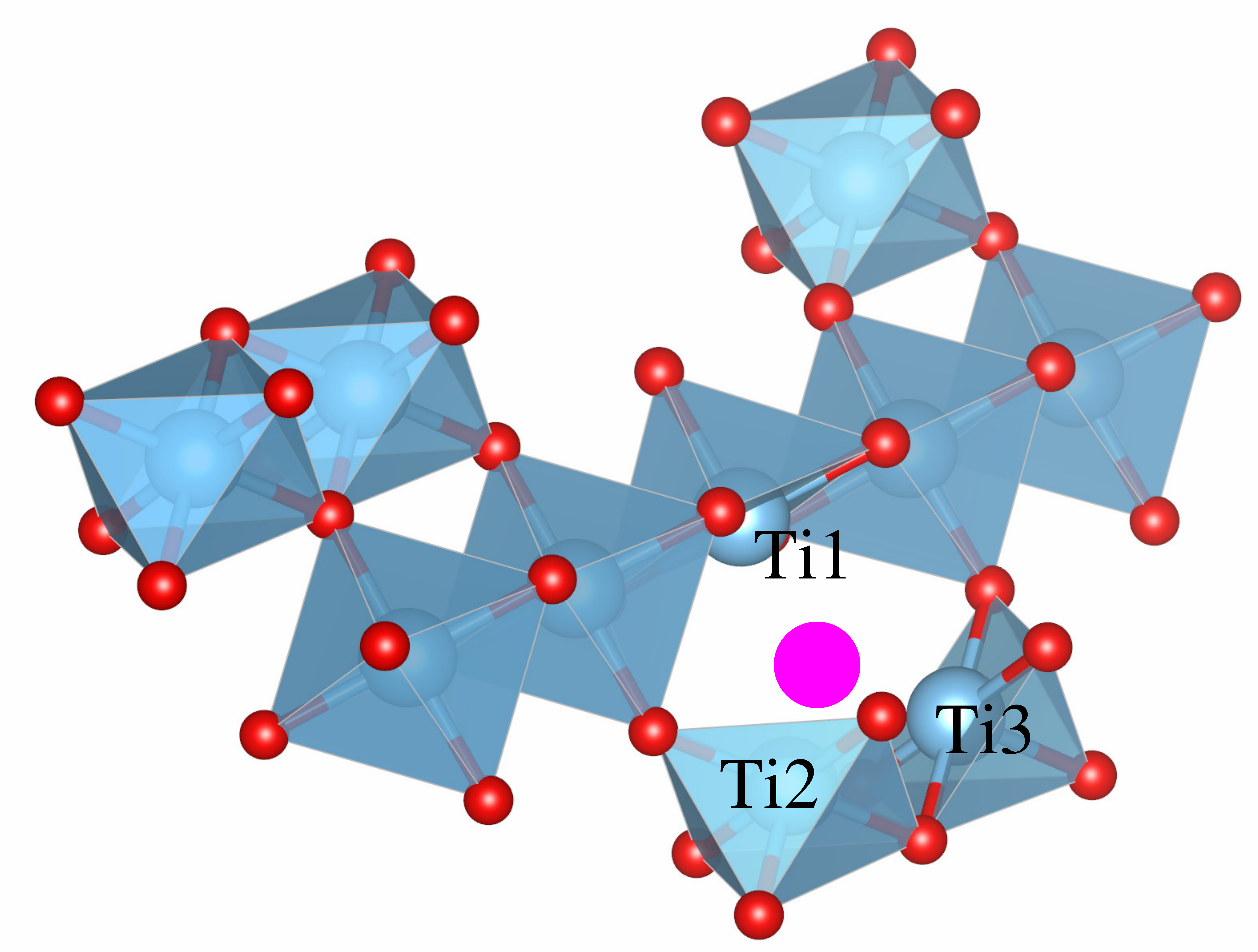}\\[0.2cm]
\caption{(color online) Structural characterization of an oxygen vacancy 
(large purple/lightgrey) in rutile.
}\label{fig:singox}
\end{figure}

As shown in 
Fig.~\ref{fig:singox}, a supercell five times the size of the primitive cell, i.e. with 10 Ti 
and 20 O atoms, serves as basis structure for the defect study. A single V$_{\rm O}$ leaves three 
nearest-neighbor titanium ions behind, here labelled Ti1, Ti2 and Ti3. In the stoichiometric
rutile structure, the Ti2-Ti3 distance marks the short side of the given Ti triangle. The 
nominal V$_{\rm O}$ concentration in this constellation amounts to $c\equiv\delta/2=0.05$, i.e. our 
modeling describes a TiO$_{1.9}$ defect case. This represents a large V$_{\rm O}$ concentration, 
however not unrealistic for the given system.~\cite{par90} Table~\ref{tab:dist} shows that the 
inter-atomic distances are only weakly modified upon structural relaxation, which may be also related
to the small supercell size. Still, the obtained pattern describes a shortening of the
Ti1-Ti2,3 distances and an elongation of the Ti2-Ti3 bonding, providing a trend to balance
the triangle distances trough the V$_{\rm O}$.

In the following we want to investigate the effect of an V$_{\rm O}$ in the rutile structure in 
terms of the local-orbital configuration as well as the net electronic structure. We will discuss 
two different choices for the correlated subspace. First in section~\ref{csub:can}, that space
is formed by {\sl all} Ti sites in the given structure, which marks the canonical and 
ground-state-oriented case. Second in section~\ref{csub:ex}, the correlated subspace is 
{\sl restricted} to the contribution of the Ti1-3 sites, i.e. it becomes more local. This second 
choice may be interpreted as treating an excited state of the system, where electrons do not see 
the explicit Coulomb repulsion on the remaining Ti sites. Since the $3d$ occupation on the original 
Ti($d^0$) sites distant to the V$_{\rm O}$ is expected small, double occupation is there rather 
rare. If moreover the electrons have gained energy from an excitation process, they can even 
more easily escape from such double occupations (as well as more efficiently screen the Coulomb 
penalties). Thus the {\sl average} effect of local $U$ and $J_{\rm H}$ on the V$_{\rm O}$-distant 
Ti sites can then be neglected to a good approximation to obtain a rough picture of the present 
system on a {\sl globally-higher} excitation level. This approximation and interpretation is also 
not in conflict with the definition of a one-particle spectral function. 
We thus simply term that space 'excited correlated subspace', and this second choice allows us to 
shed light on possible changes in the oxidation state of the oxygen vacancy. Still note however
that the Coulomb interactions on the Ti sites away from the V$_{\rm O}$s are crucial to understand
the semiconducting character of TiO$_{2-\delta}$, as will be explained in the following 
section~\ref{csub:can}.
\begin{table}[t]
\begin{ruledtabular}
\begin{tabular}{l|ccc}
               &  Ti1$-$Ti2  & Ti1$-$Ti3  & Ti2$-$Ti3   \\ \hline
TiO$_2$        &  3.57     &  3.57    & 2.96     \\
TiO$_{1.9}$    &  3.53     &  3.53    & 2.97     \\
\end{tabular}
\end{ruledtabular}
\caption{Comparison of the inter-atomic distances (in \AA) between Ti ions 
surroundig a vacancy-designated oxygen site.}
\label{tab:dist}
\end{table}

\subsubsection{Canonical correlated subspace formed by all Ti sites\label{csub:can}}
\begin{figure}[t]
\includegraphics*[width=8.25cm]{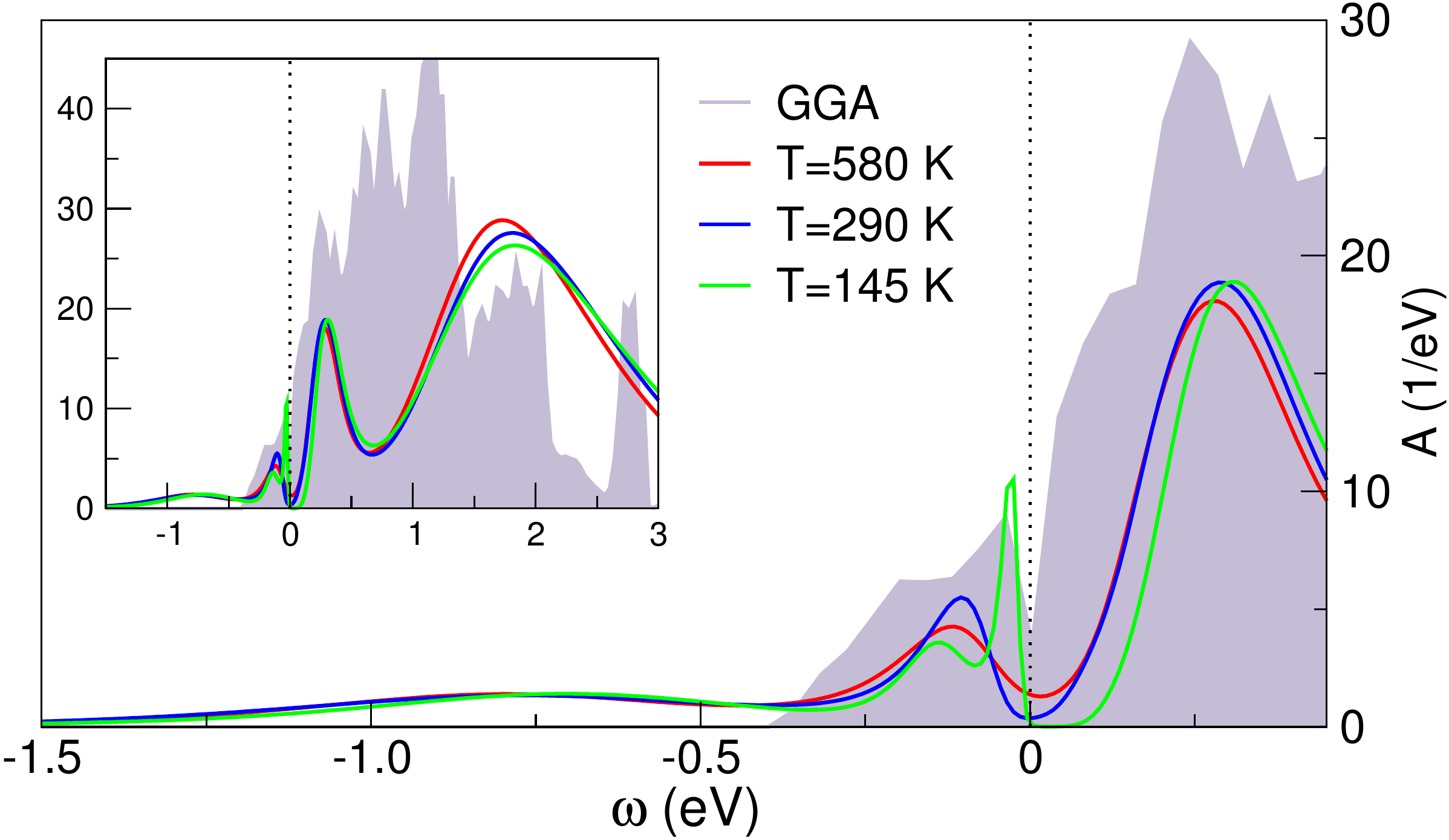}
\caption{(color online) Total spectral information for rutile-TiO$_{1.9}$ in GGA and DFT+DMFT.
at different temperatures in two energy windows.}\label{fig:elox}
\end{figure}
Let us start with the well-defined DFT+DMFT setting of the correlated subspace build by all
the Ti sites. On the GGA level, as displayed by plotting the total spectral function 
$A(\omega)=\sum_{\bf k} A({\bf k},\omega)$ in Fig.~\ref{fig:elox}, the considered system
becomes metallic with semimetallic tendencies around the Fermi level $\varepsilon_{\rm F}$ 
positioned within the $t_{2g}$ manifold. Yet correlations and finite temperature $T$ render 
the situation more intriguing. At low $T$ the defect structure is semiconducting within 
DFT+DMFT, with a small charge gap $\Delta\sim 0.06$\,eV at $T=145$\,K. That gap is filled with 
rising temperature, marking a bad-metal regime. In addition, there is sizable transfer of 
spectral weight to a broad in-gap structure centered at 
$\varepsilon_{\rm IG}^{\hfill}\sim -0.75$\,eV. Note that optics measurements detect an absorption 
peak in reduced semiconducting TiO$_2$ single crystals at 0.75\,eV,~\cite{cro51,cro52} commonly
used to explain its blue color.~\cite{par90} Photoemission measurements on the rutile surface 
report a defect-state
peak at $\sim -0.9$\,eV,~\cite{tho07,wen08,yim10} while on the anatase surface it is located 
at an higher energy of $\sim -1.1$\,eV~\cite{tho07,roe15} Furthermore, scanning tunneling 
spectroscopy finds an in-gap state at $\sim -0.7$\,eV on the defect-rutile surface~\cite{set14}
and x-ray photoelectron spectroscopy on rutile TiO$_2$ nanoparticles reveals a defect state
at $\sim -0.8$\,eV.~\cite{vas16} 
It is tempting to relate these experimental findings of deep levels in TiO$_{2-\delta}$ to the 
present satellite peak. On the other hand, $n$-type conductivity with rather high mobilities
due to shallow level has been also reported in the literature,~\cite{yag96} which might be 
connected to our small-gap feature. 

To gain insight in the nature and characteristics of the Ti-local states near the V$_{\rm O}$,
first Tab.~\ref{tab:orbs} provides the effective orbitals on Ti1-3 written in terms of linear 
combinations of the original $(t_{2g},e_g)$ functions from Tab.~\ref{tab:rut}. The $\varphi$
and $\varphi'$ orbitals on Ti2,3 behave very similarly, therefore we restrict the discussion
to the $\varphi$ branch. It is seen that 
while the $\psi_2$ orbital has strong $e_g$ character, the orbitals $\varphi_1$, $\varphi_2$ 
have sizable contributions from both original orbital sectors. Thereby $\varphi_1$ is 
$t_{2g}$-dominated and $\varphi_2$ is $e_{g}$-dominated. Thus a nearly exclusive $e_g$ character 
of the local defect states, as e.g. given in oxygen-deficient SrTiO$_3$,~\cite{luo04,lec16} 
does not apply for oxygen vacancies in TiO$_2$. Note furthermore that the $e_g'(2)$ orbital has 
almost negligible contribution to the V$_{\rm O}$-induced physics, as the $e_g'(2)$-dominated 
$\varphi_3$ orbital remains nearly empty.

Concerning tight-binding parameters, the hopping between both $\varphi_1$ on Ti2 and Ti3 is 
largest with $t_{\varphi_1}=-0.2$\,eV, while the other hopping amplitudes on the Ti triangle 
are of absolute value $\le 0.1$\,eV. Non-surprisingly, this marks Ti2,3 as more strongly 
coupled, thus possibly prone to singlet/triplet formation.

\begin{figure}[t]
\includegraphics*[width=8.25cm]{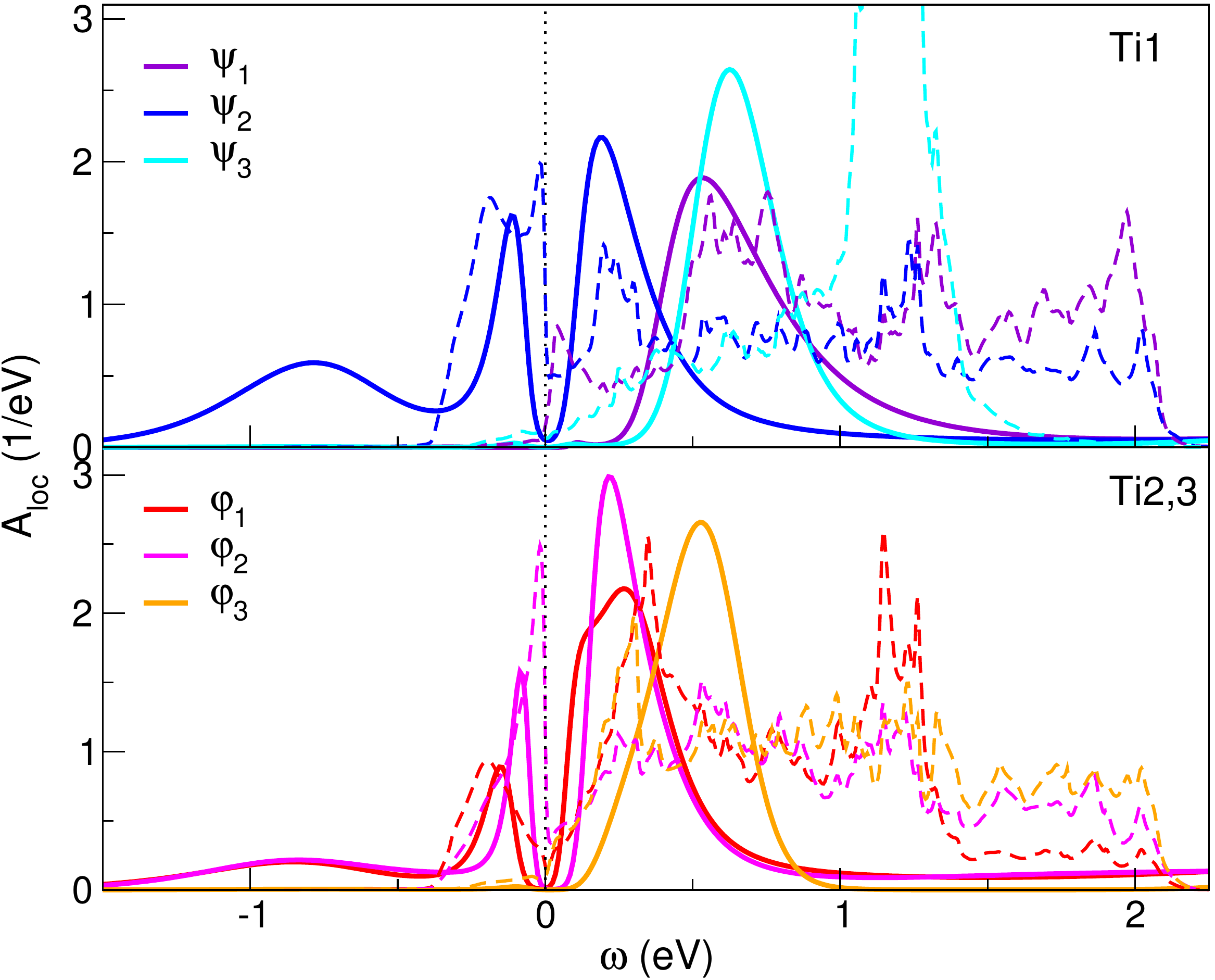}
\caption{(color online) Local spectral information for rutile-TiO$_{1.9}$ on Ti1 and 
Ti2,3 (cf. Fig.~\ref{fig:singox}) at $T=290$\,K. Dashed lines correspond to the GGA result, 
respectively.}\label{fig:locox}
\end{figure}

Figure~\ref{fig:locox} provides the spectra on Ti1-3, which are expected to display (partly) 
Ti$^{3+}$ character. Note that although all Ti sites in the supercell contribute to the full 
correlated subspace, electron occupation on the remaining Ti sites is very small. The filling on 
the Ti sites farther away from V$_{\rm O}$ is also not significantly raised with temperature.
The effective orbitals on Ti2 and Ti3 are nearly equivalent by symmetry and
behave here very similarly, so no site differentiation is needed. Of the three correlated 
orbitals $\{\psi_m\}$ on Ti1, only $\psi_2$ has sizable filling. 
\begin{table}[t]
\begin{ruledtabular}
\begin{tabular}{lc|rrrrrr}
site & orbital   & $\varepsilon^{\hfill}_{\rm CF}$ & $|e_g'(2)\rangle$ & $|e_g'(1)\rangle$ & $|a_{1g}\rangle$ & $|e_g(1)\rangle$ 
& $|e_g(2)\rangle$  \\ \hline
     & $|\psi_1\rangle$      & 1102 & 0.035 &-0.999 & 0.000 &   0.000 & -0.002 \\
Ti1  & $|\psi_2\rangle$      &  698 & 0.000 & 0.000 &-0.147 &  -0.006 &  0.990 \\
     & $|\psi_3\rangle$      & 1023 & 0.000 & 0.001 & 0.990 &  -0.003 &  0.147 \\[0.2cm]
     & $|\varphi_1\rangle$   &  710 &-0.032 &-0.023 & 0.837 &  -0.427 & -0.214 \\
Ti2  & $|\varphi_2\rangle$   &  785 & 0.077 & 0.073 & 0.308 &   0.733 &  0.473 \\
     & $|\varphi_3\rangle$   &  992 &-0.759 &-0.642 &-0.221 &   0.096 &  0.058 \\[0.2cm]
     & $|\varphi_1'\rangle$  &  709 & 0.036 &-0.027 & 0.872 &   0.437 & -0.220 \\
Ti3  & $|\varphi_2'\rangle$  &  780 & 0.083 &-0.079 &-0.488 &   0.726 & -0.470 \\
     & $|\varphi_3'\rangle$  &  992 & 0.758 &-0.641 & 0.018 &  -0.103 &  0.063 \\
\end{tabular}
\end{ruledtabular}
\caption{Effective V$_{\rm O}$-induced orbitals on the three nearby Ti sites in the rutile
structure, with their respective crystal-field level $\varepsilon^{\hfill}_{\rm CF}$ (in meV), 
written in terms of linear combinations of the original $(t_{2g},e_g)$ functions.}
\label{tab:orbs}
\end{table}
Spurious occupation of $\psi_1,\psi_3$ in GGA is eliminated by correlations. On the contrary, 
two orbitals, i.e. $\varphi_1$ and $\varphi_2$ are occupied on Ti2,3. The designation of $\psi_2$,
$\varphi_1$ and $\varphi_2$ is already suggested from their favorable crystal-field levels
(cf. Tab.~\ref{tab:orbs}). In DFT+DMFT the occupations 
read $(n_{\psi_2},\,n_{\varphi_1},\,n_{\varphi_2})=(0.65,\,0.29,\,0.33)$, thus all three Ti sites 
have similar occupation, however $n_{\rm Ti1}>n_{\rm Ti2,3}$ holds. Since the site-resolved
fillings sum up to $n_{V_{\rm O}}=1.89$ and oxygen is in the O$^{2-}$ oxidation state in the
compound, the defect is described as of being close to the nominal neutral vacancy state 
V$_{\rm O}^0$.

\begin{figure}[t]
\hspace*{-0.1cm}\includegraphics*[width=8.25cm]{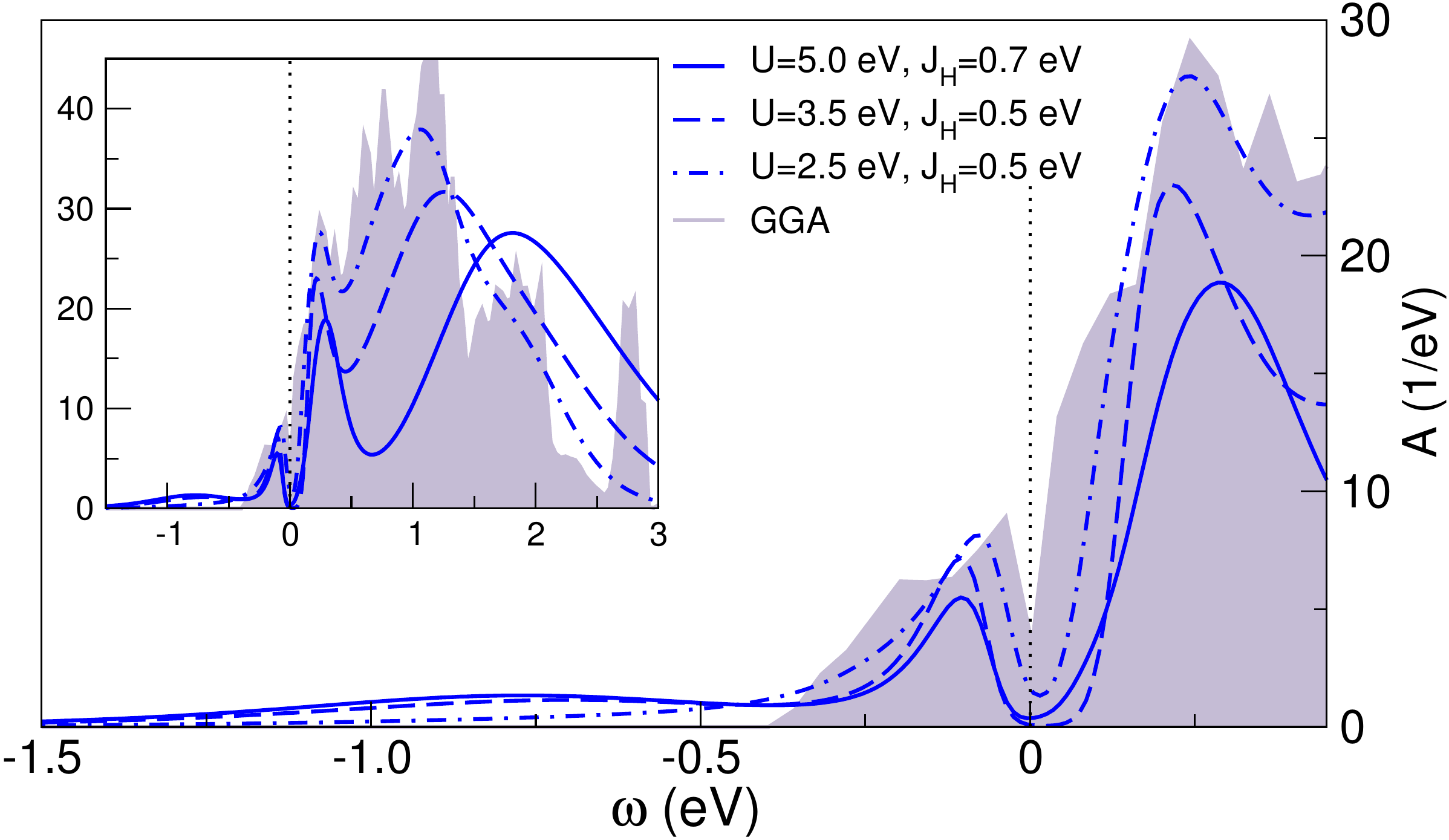}\\[0.1cm]
\caption{(color online) Total spectral function for rutile TiO$_{1.9}$ based on three different 
sets of local Coulomb interactions at $T=290$\,K.}\label{fig:eloxu}
\end{figure}

In order to check for the influence of the chosen local Coulomb interactions on the physics,
we additionally peformed calculations for $U=3.5$\,eV and $U=2.5$\,eV, both with 
$J_{\rm H}=0.5$\,eV. The values $U=3.5$\,eV, $J_{\rm H}=0.5$\,eV were used in a recent 
DFT+DMFT study of the oxygen-deficient SrTiO$_3$ surface.~\cite{lec16} There are no
qualitative differences between the resulting total spectral functions (cf. Fig.~\ref{fig:eloxu})
for the sets ($U=5$\,eV, $J_{\rm H}=0.7$\,eV) and ($U=3.5$\,eV, $J_{\rm H}=0.5$\,eV), hence the 
detected physics is rather stable within a reasonable range of local Coulomb-interaction 
parameters. Only at much smaller $U=2.5$\,eV the semiconducting gap as well as the deeper
in-gap state seem to disappear.

It remains to specify the likely mechanism behind the DFT+DMFT finding. The results suggest that the
V$_{\rm O}$ provides electron doping by initially forming rather shallow states below the
original $t_{2g}$ mainfold. Because of the higher O-Ti connectivity and the entangled
($e_g$,$t_{2g}$) defect signature in rutile, both compared to e.g. surface SrTiO$_3$,~\cite{lec16} 
the coherency of those states is increased and they seemingly develop band-like character. Sizable 
Coulomb interactions then lead to a Mott criticality, resulting in band renormalization and 
formation of Hubbard(-like) bands. The latter give rise to the 
deep-level in-gap spectra. Importantly, the driving force between the Mott(-like) gap formation 
is different from a conventional correlated multi-band lattice problem. Since the local states 
near the V$_{\rm O}$ are connected {\sl between} different V$_{\rm O}$s by fragile hopping paths, 
mainly the local Coulomb interactions on the Ti sites {\sl distant} from the defects are determinative 
for driving the doped system again insulating. In other words, the 'interstitial' Coulomb-repulsive 
region between the V$_{\rm O}$s destroys the fragile band-like defect states and localizes the 
electrons dominantly near the oxygen vacancy.

\subsubsection{Excited correlated subspace formed by Ti1, Ti2, Ti3\label{csub:ex}}
\begin{figure}[b]
(a)\hspace*{-0.1cm}\includegraphics*[width=8cm]{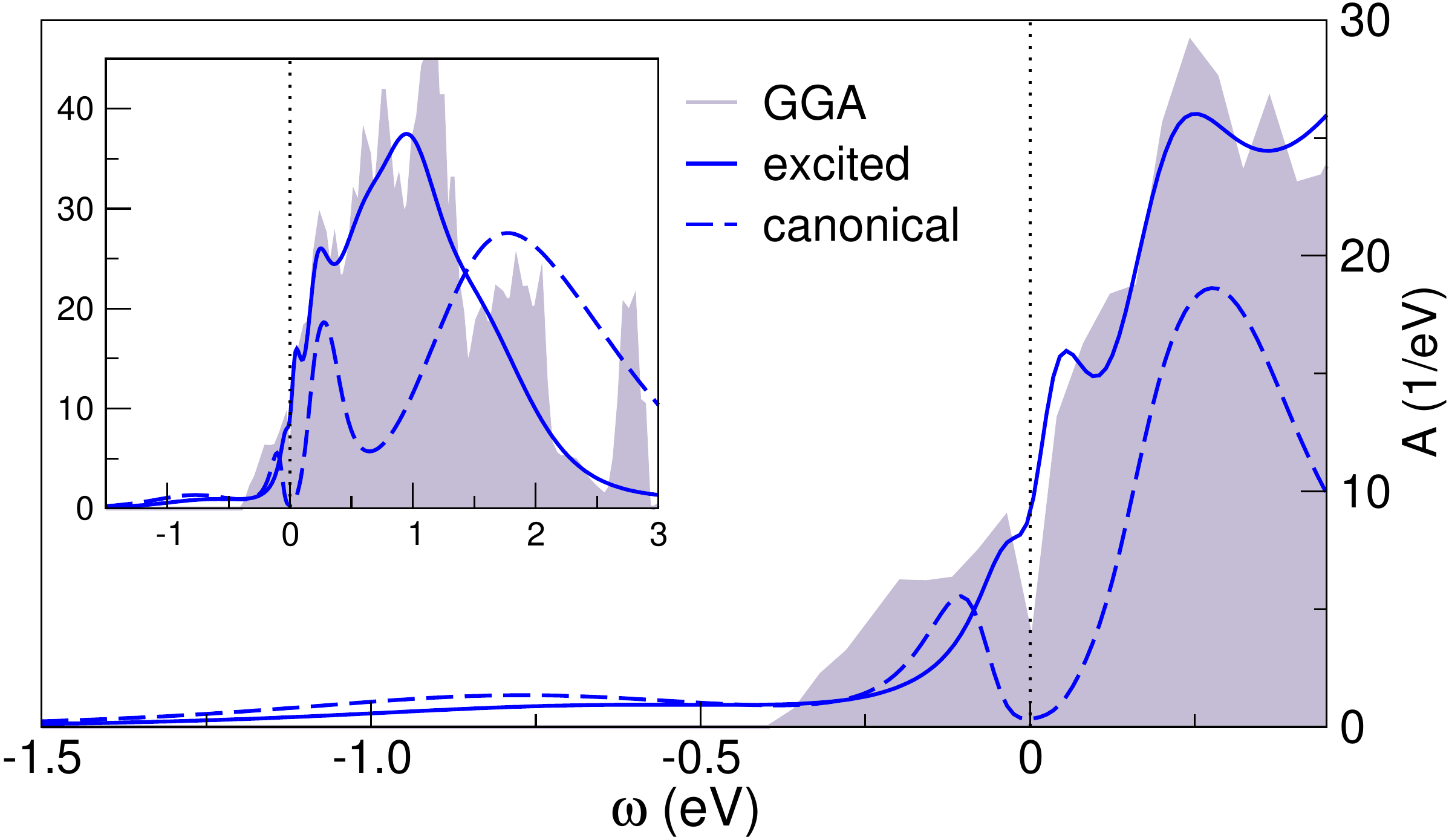}\\[0.1cm]
(b)\hspace*{-0.2cm}\includegraphics*[width=8cm]{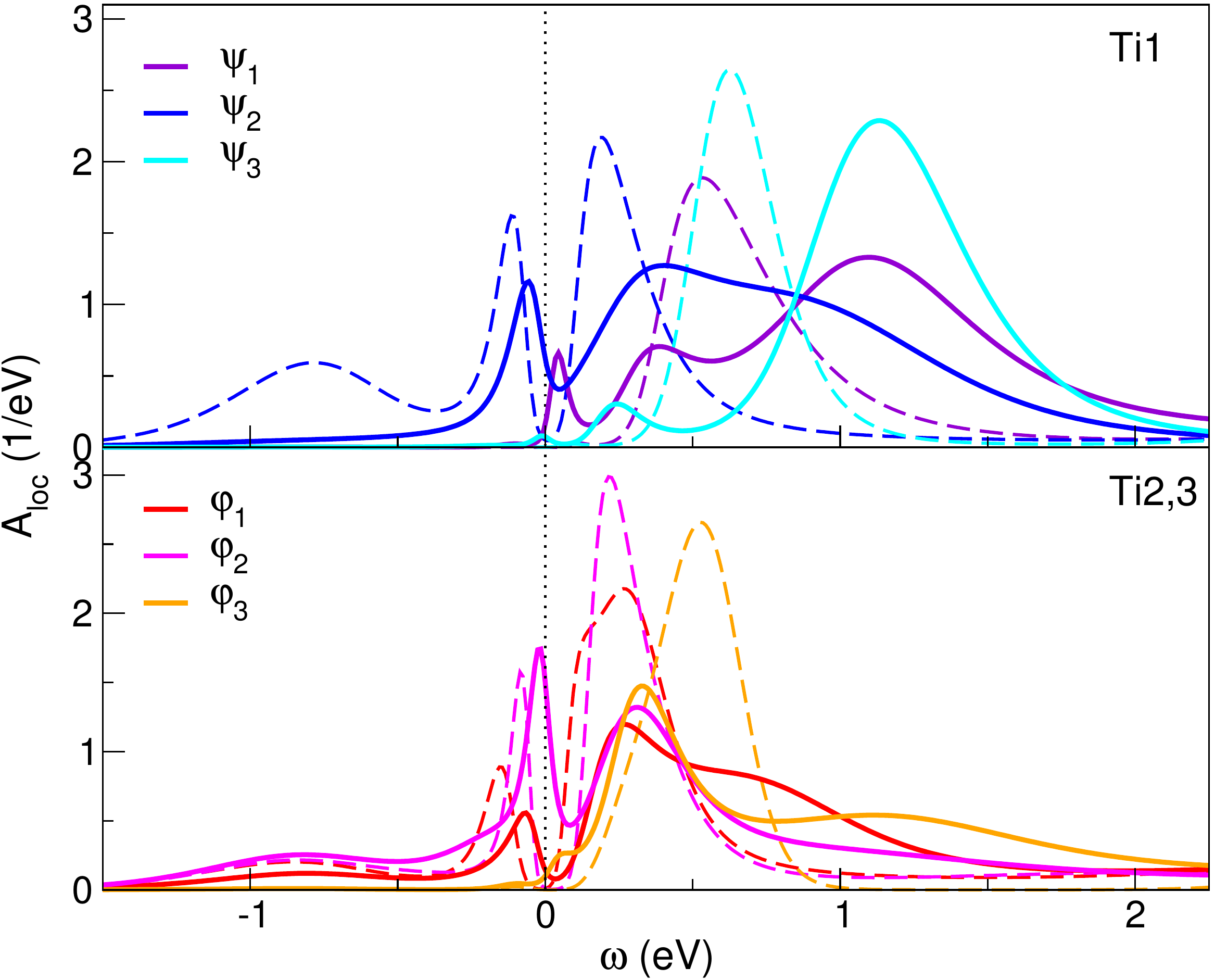}
\caption{(color online) Comparison of spectral information for rutile-TiO$_{1.9}$ using
the canonical and the excited correlated subspace at $T=290$\,K.
(a) Total spectral function. 
(b) Local spectra for Ti1 and Ti2,3 (cf. Fig.~\ref{fig:singox}) at $T=290$\,K,
full lines: excited, dashed lines: canonical. 
}\label{fig:eloxex}
\end{figure}
Letting only the sites Ti1-3 contribute to the correlated subspace serves to goals. First, it 
gives access to a possibly different V$_{\rm O}$ charging state. Second, it provides a check for 
our presented mechanism, denoting the Coulomb interactions on Ti away from V$_{\rm O}$ as being 
mainly to be blamed for the semiconducting character. 

In general, the states V$_{\rm O}^0$, V$_{\rm O}^{+1}$ and V$_{\rm O}^{+2}$ are discussed for 
oxygen-deficient TiO$_2$,~\cite{yan09,mat10,jan13,lin15} which in other works appeared relevant 
to fit findings of shallow-donor properties and well-localized defect 
states within a coherent picture. As discussed in section~\ref{csub:can}, the present DFT+DMFT study 
already provides means to such a coexistence by revealing small-gap {\sl as well as} deeper 
in-gap features. Nonetheless, by performing additional calculations within the local-restriced
correlated subspace one may learn further details of the V$_{\rm O}$ energy-level structure.
Note that in principle the local Coulomb interactions also change when the 
correlated subspace is modified, i.e. in the present scenario should be lowered for the smaller
subspace. But for simplicity we keep $U=5$\,eV and $J_{\rm H}=0.7$\,eV on Ti1-3 to reveal the 
key features of the excited system.

As expected, a metallic solution results from the calculation, with reduced and shifted
incoherent weight at higher energy (see Fig.~\ref{fig:eloxex}a). This is indeed in favor of
our 'interstitial Coulomb' mechanism being relevant for fully localizing electrons near
V$_{\rm O}$s. 
Interestingly, the Ti1-3 sites now display a rather different local filling in the dominant 
orbitals, namely $(n_{\psi_2},\,n_{\varphi_1},\,n_{\varphi_2})=(0.24,\,0.19,\,0.43)$, and a 
total filling of these sites amounting to $n_{V_{\rm O}}=1.56$. This means that in the excited 
scheme not only the total filling of Ti close to V$_{\rm O}$ is substantially reduced compared to
the canonical scheme, but also the local filling symmetry is different. Now the Ti2,3
sites are majorly occupied and the Ti1 site plays the weaker role, also visualized by
plotting the local spectral functions in Fig.~\ref{fig:eloxex}b. The high-energy, incoherent
and strongly-localized part of the spectra is now dominantly carried by states on Ti2,3.
It appears that in the excited-system calculation, localized charge on Ti1 has been partially
transfered to itinerant states.

These findings are indeed reminiscent of the identification of different V$_{\rm O}$ charging 
states in DFT+U and/or DFT hybrid-functional studies.~\cite{mat10,jan13,lin15} Also the
characterization of two different site-orbital levels has been reported before, yet usually
by invoking explicit magnetic ordering. Our two detected charging states with 
$n^{\rm can}_{V_{\rm O}}=1.89$, reading V$_{\rm O}^{+0.11}$, and $n^{\rm ex}_{V_{\rm O}}=1.56$, 
reading V$_{\rm O}^{+0.44}$, do not very strongly deviate from the neutral-vacancy case. 
However note that highly-oxidized states like V$_{\rm O}^{+1}$/V$_{\rm O}^{+2}$ are usually 
found in the presence of additional trivalent substitutional impurities on the Ti site, such as 
Fe$^{3+}$ and Cr$^{3+}$.~\cite{yan09} 
An established theoretical picture~\cite{jan13} describes the vacancy-defect state as a
bound object consisting of V$^{2+}_{\rm O}$ plus two polarons, rendering it charge
neutral again. In principle one may try to interprete our results also along such lines, 
since the present filling scenario with Hubbard-like high-energy spectral parts point 
to a local $S=1$ spin in the paramagnetic material. On the other hand, as discussed, we do 
not {\sl have to} invoke the polaron picture on clear grounds to account for the coexistence of
shallow and deep states.

Let us finally note that the overall qualitative picture that we here derived for a single 
V$_{\rm O}$ in rutile TiO$_2$ is believed to be stable against modified structural relaxations, 
as e.g. provided by a different(larger) supercell. More dilute V$_{\rm O}$ cases could still lead 
to a weakening of the shallow levels in favor of the deep level, with an increase of the 
semiconducting gap. Within the local-level manifold the subtle details of the energy hierachy 
might be affected by structural issues.  But the general notion of the relevant $\psi_2$ on Ti1 
as well as $\varphi_1$ and $\varphi_2$ on Ti2,3 appears robust.

\subsection{Higher oxygen-vacancy concentration: Ti$_3$O$_5$}
\subsubsection{Spectral properties and total energies}
\begin{figure}[b]
(a)\includegraphics*[width=8cm]{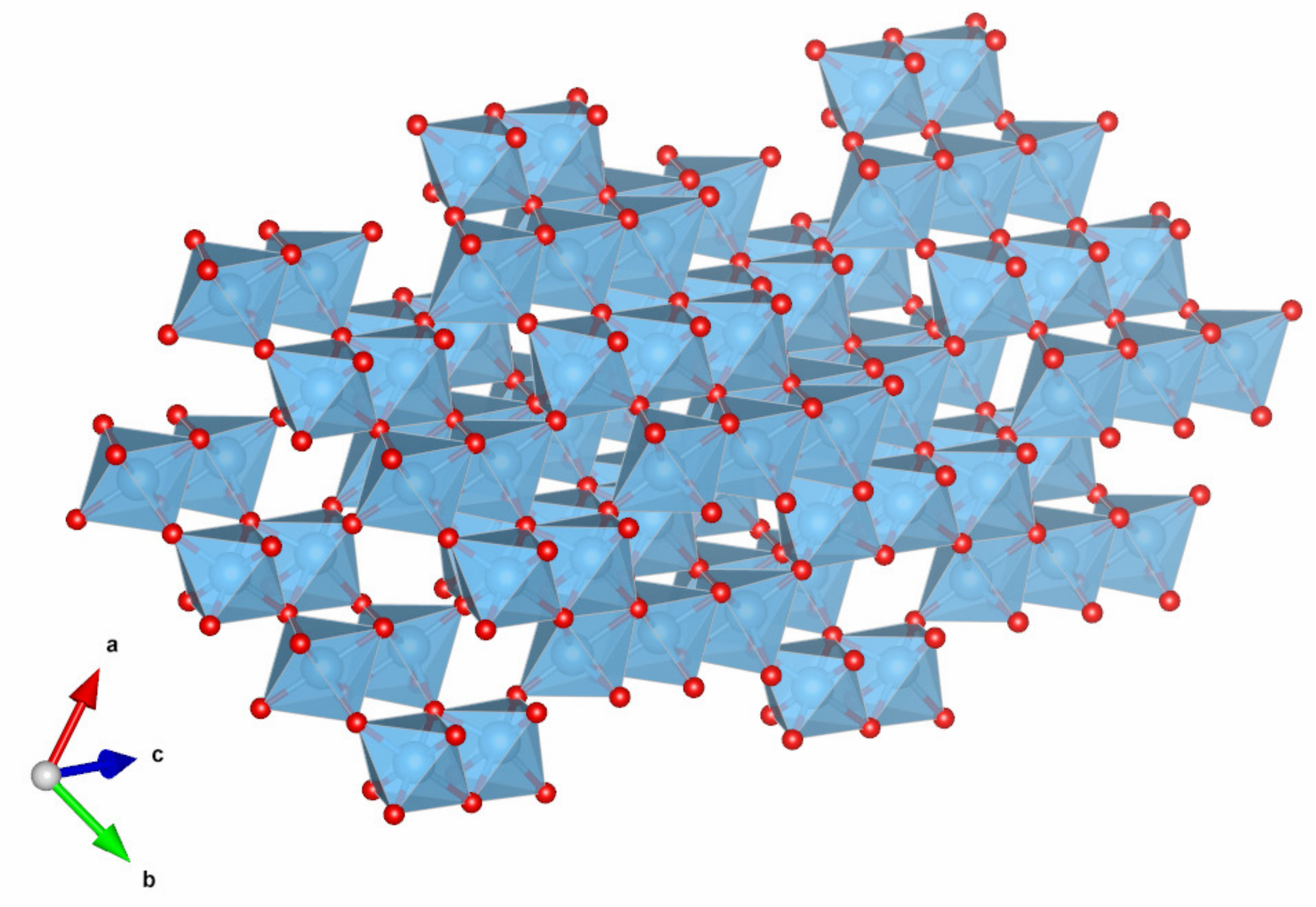}\\
(b)\includegraphics*[width=8cm]{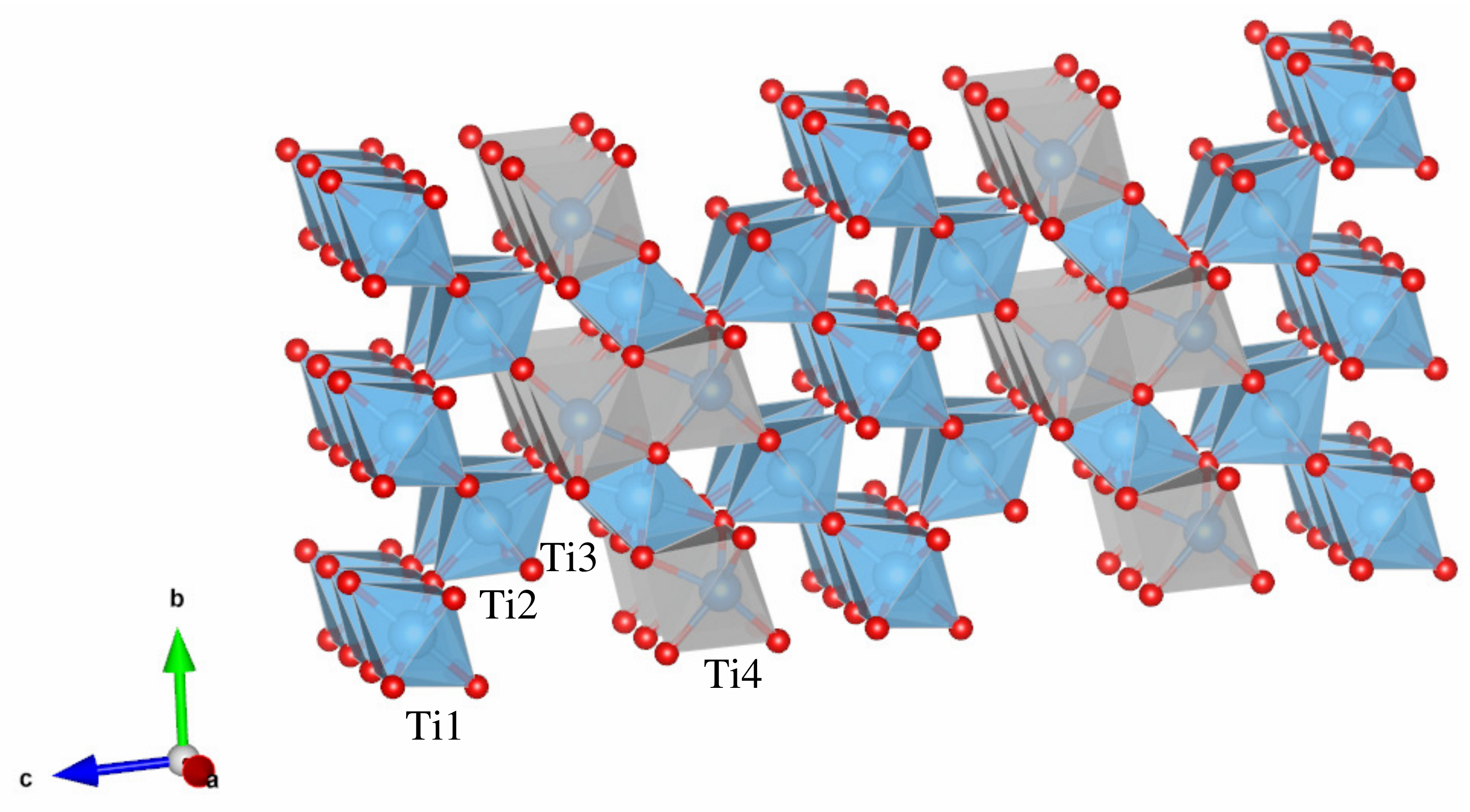}
\caption{(color online) Illustration of (a) the $\gamma$-Ti$_3$O$_5$ structure 
and (b) the defect-rutile Ti$_3$O$_5$ structure, both with
Ti (large blue/grey) and O (small red/dark) atoms. In (b), the labels Ti1-4 mark 
symmetry-inequivalent Ti sites, with grey polyhedra surrounding the
5-fold coordinated Ti4 sites.}\label{fig:ti3o5}
\end{figure}
We now shift attention to the problem of oxygen vacancies at higher concentration.
In order to gain insight in the properties of V$_{\rm O}$s in a designated 
ordered limit, we study a specific Magn{\'e}li compound at Ti$_3$O$_5$ stoichiometry.
There is strong interest in the various Magn{\'e}li phases at 
Ti$_3$O$_5$,~\cite{asb59,hon82,ono98,ohk10,pad14,tan15} since e.g. photoreversible phase 
transitions occur at room temperature. But as we are interested in the main effect of 
vacancy ordering, we focus on a single allotrop, the so-called $\gamma$ phase. 
The underlying crystal structure has monoclinic symmetry and can be stabilized at room 
temperature.~\cite{hon82} 

Described in simple terms, whereas perovskite SrTiO$_3$ has corner-sharing TiO$_6$ octhedra 
and rutile TiO$_{2}$ an elementary alternation of corner- and face-sharing octahedra, 
the Magn{\'e}li phases exhibit more complicated arrangement of those two octahedra-sharing 
types to accomodate a desired stoichiometry (see Fig.~\ref{fig:ti3o5}a). This may be 
interpreted as an ordering of vacancies, however importantly, contrary to e.g. our 
TiO$_{1.9}$ structure, there is no V$_{\rm O}$-induced 'destruction' of local TiO$_6$ 
octhedra. In terms of a formal oxygen deficiency $\delta$, the given Ti$_3$O$_5$ stoichiometry 
amounts nonetheless to $\delta=0.33$, i.e. the compound would correspond to TiO$_{1.67}$ with
vacancy concentration $c=0.167$.

In order to compare the electronic characteristics of the optimal-ordered Magn{\'e}li 
structure with V$_{\rm O}$s in rutile, we in addition performed calculations for 
defect rutile with stoichiometrty Ti$_3$O$_5$. The corresponding crystal structure 
(cf. Fig.~\ref{fig:ti3o5}b) was determined by energy minimization of oxygen-vacancy 
arrangements at the desired composition by making use of the cluster-expansion 
technique (see Ref.~\onlinecite{hec15} for details). In this defect-rutile structure,
part of the TiO$_6$ are indeed 'damaged', resulting in selected fivefold-O coordinated 
Ti sites,here designated as Ti4.
A total-energy comparison on the GGA level, as expected, clearly favors the Magn{\'e}li
structure. The more clever restoration of ideal 6-fold-oxygen coordination around Ti 
is appreciated by a substantial $\sim 0.42\,$eV per Ti atom against defect rutile.

\begin{figure}[t]
\hspace*{-0.1cm}\includegraphics*[width=8.5cm]{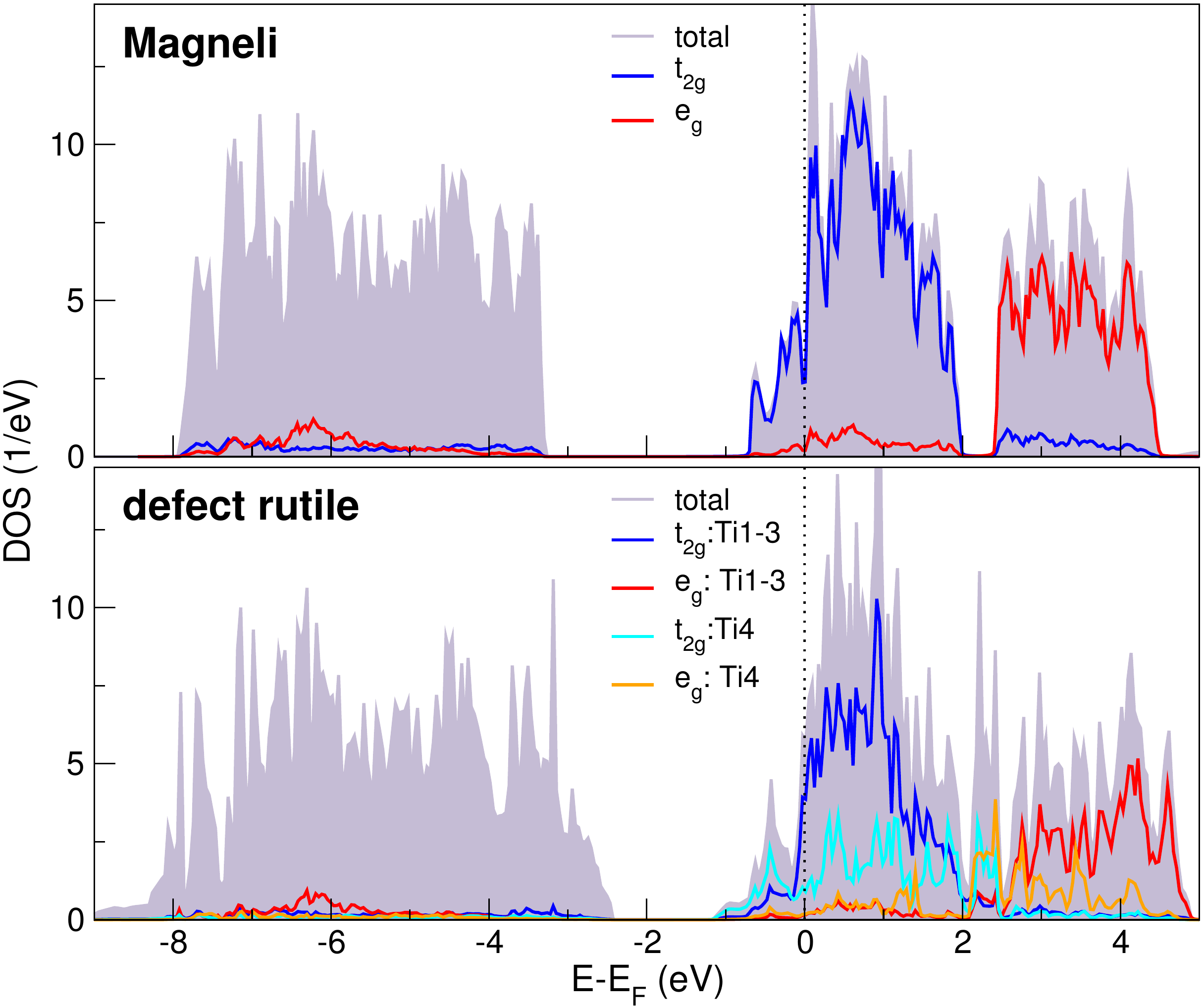}
\caption{(color online) Comparison of the GGA density of states of 
$\gamma$-Ti$_3$O$_5$ (top) and defect-rutile Ti$_3$O$_5$ (bottom), in view of 
($t_{2g}$,$e_g$) contributions. The labeling Ti1-3 and Ti4 referes to the sites marked
in Fig.~\ref{fig:ti3o5}b.}\label{fig:ti3o5-dft}
\end{figure}

Figure~\ref{fig:ti3o5-dft} documents the differences in the GGA density of states in terms
of the partitioning in Ti($t_{2g}$,$e_g$)-like contributions. Based on the robust 
TiO$_6$-octahedral structuring of the Magn{\'e}li phase, the electronic structure of
$\gamma$-Ti$_3$O$_5$ shows a clear distinction into those $3d$ submanifolds. This is partly
also true for defect-rutile Ti$_3$O$_5$, yet the $(t_{2g},e_g)$ states from the
TiO$_5$ polyhedra show there more substantial overlap in energy. In fact, the small
$t_{2g}$-$e_g$ gap region above 2\,eV in $\gamma$-Ti$_3$O$_5$ just becomes filled by
such states in the defect-rutile structure.

\begin{figure}[t]
\hspace*{-0.1cm}\includegraphics*[width=8.5cm]{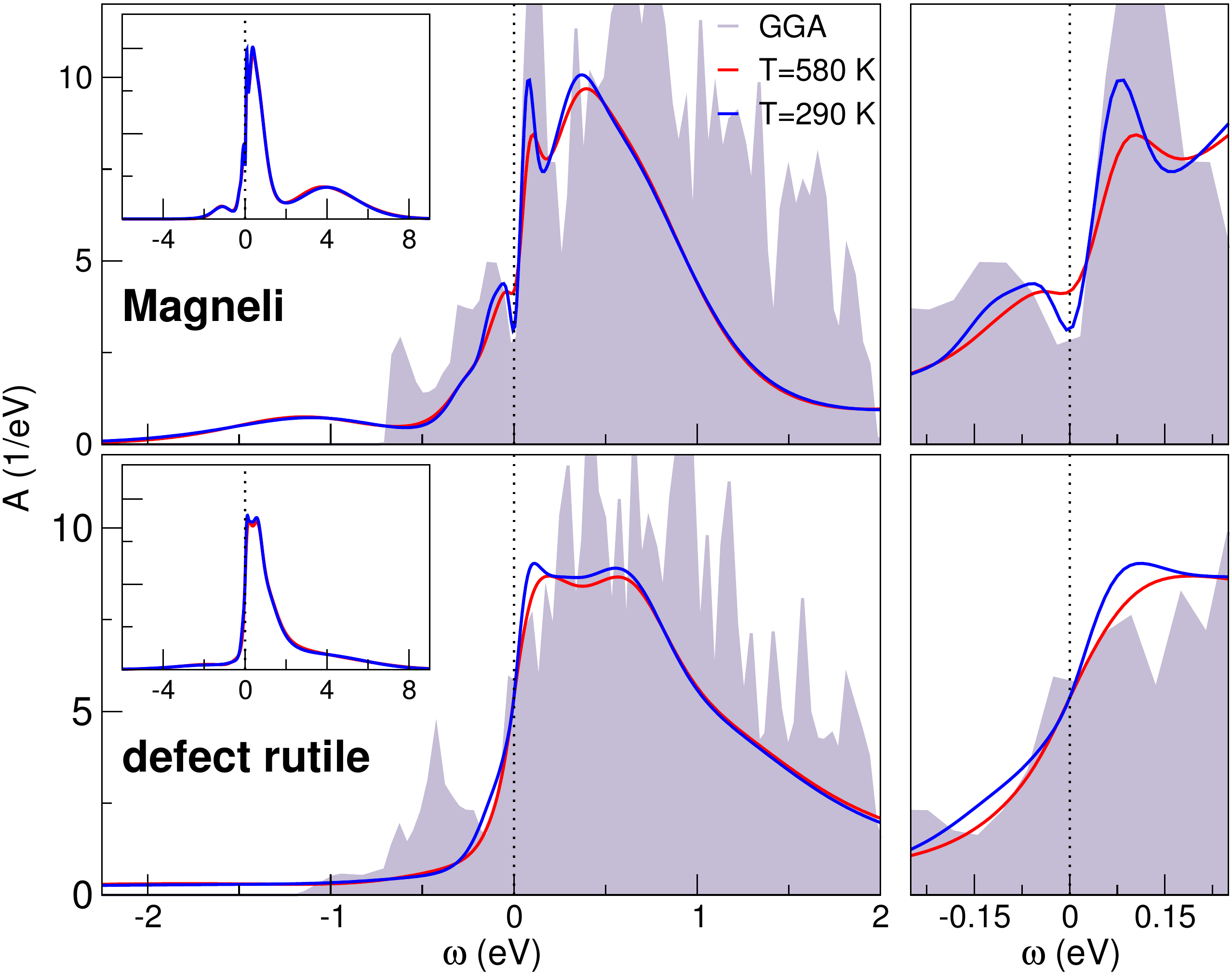}
\caption{(color online) Comparison of the correlated total spectral function of 
$\gamma$-Ti$_3$O$_5$ (top) and defect-rutile Ti$_3$O$_5$ (bottom) at two different
temperatures.}\label{fig:comp}
\end{figure}

The total spectral functions obtained within DFT+DMFT are plotted in Fig.~\ref{fig:comp}.
Both systems remain metallic with including many-body correlations. A lower Hubbard band 
located at $\sim -1.1\,$eV and renormalization at low energy are identified for the 
Magn{\'e}li phase. This marks the system as a seemingly 'textbook' correlated material,
with coherent renormalized quasiparticles at low energy and incoherent Hubbard bands at
higher energies. Already on the GGA level there is a small spectral dip at the Fermi level, 
verified also with correlations at lower temperature. Note that the Hubbard band is of course 
nearly exclusively formed by $t_{2g}$-like states. Satellite structures of $t_{2g}$ kind
have recently been detected in a hard x-ray photoelectron spectroscopy study of the structurally 
different $\beta$-Ti$_3$O$_5$ and $\lambda$-Ti$_3$O$_5$ Magn{\'e}li compounds.~\cite{kob17}

Though the same total filling scenario holds for the defect-rutile phase, the characterization 
with electronic correlations appears more subtle. First, in view of the semiconducting 
defect-rutile state discussed in section~\ref{csub:can}, the metallic response at higher 
V$_{\rm O}$ concentration is not that surprising. The fragile hopping paths between 
V$_{\rm O}$-near electronic states become increasingly robust with growing concentration
of vacancies, such that eventually Coulomb repulsion is not anymore capable of establishing
a charge-gapped material.
In the low-energy spectrum there is some renormalization due to correlations, but not of 
significant kind. On a first glance, the spectral function looks furthermore rather monotonic 
in the occupied part, yet importantly, not necessarily implying that correlations are weak. 
Because the comparison to the GGA results shows, there is significant spectral-weight 
transfer to energies far away from the Fermi level, too. However this transfer does not give 
rise to an obvious lower Hubbard-band peak, but is broadly distributed over a wider energy
range (see inset of Fig.~\ref{fig:comp}). This finding is somewhat reminiscent of
obvervations made in resonant-photoemission experiments on electron-doped 
SrTiO$_3$.~\cite{ishi08} There, broader in-gap weight was assigned to the increased 
relevance of Ti($3d$)-O($2p$) hybridization. Since the $e_g$ character, which is more 
strongly hybridized with O($2p$) than $t_{2g}$, plays a more prominent role in the 
defect-rutile case, the present result could possibly point to a similar interpretation.
On the other hand, a recent GW+DMFT study suggests the possibility of diminished
satellite peaks in some correlated compounds, as well as a reinterpretation of their 
original character.~\cite{boe16}

The total-energy difference between both Ti$_3$O$_5$ structural types is
even slightly increasing with including many-body correlations. At room temperature, the 
Magn{\'e}li phase is favored by $\sim 0.46\,$eV per Ti atom.

\subsubsection{Charge disproportionation between Ti sites}
So far we did not comment on the local occupations of the respective Ti sites in 
$\gamma$- and defect-rutile Ti$_3$O$_5$. Since the nominal oxidation state at that
stoichiometry amounts to Ti$^{3.33+}$, charge fluctuations are expected to be more 
relevant than in many other oxides with nominal integer Ti valence. In fact, charge ordering in
connection with a metal-insulator transition is commonly discussed for various Magn{\'e}li 
phases, especially for the Ti$_4$O$_7$ compound.~\cite{leo06}

\begin{figure}[t]
\hspace*{-0.1cm}\includegraphics*[width=8cm]{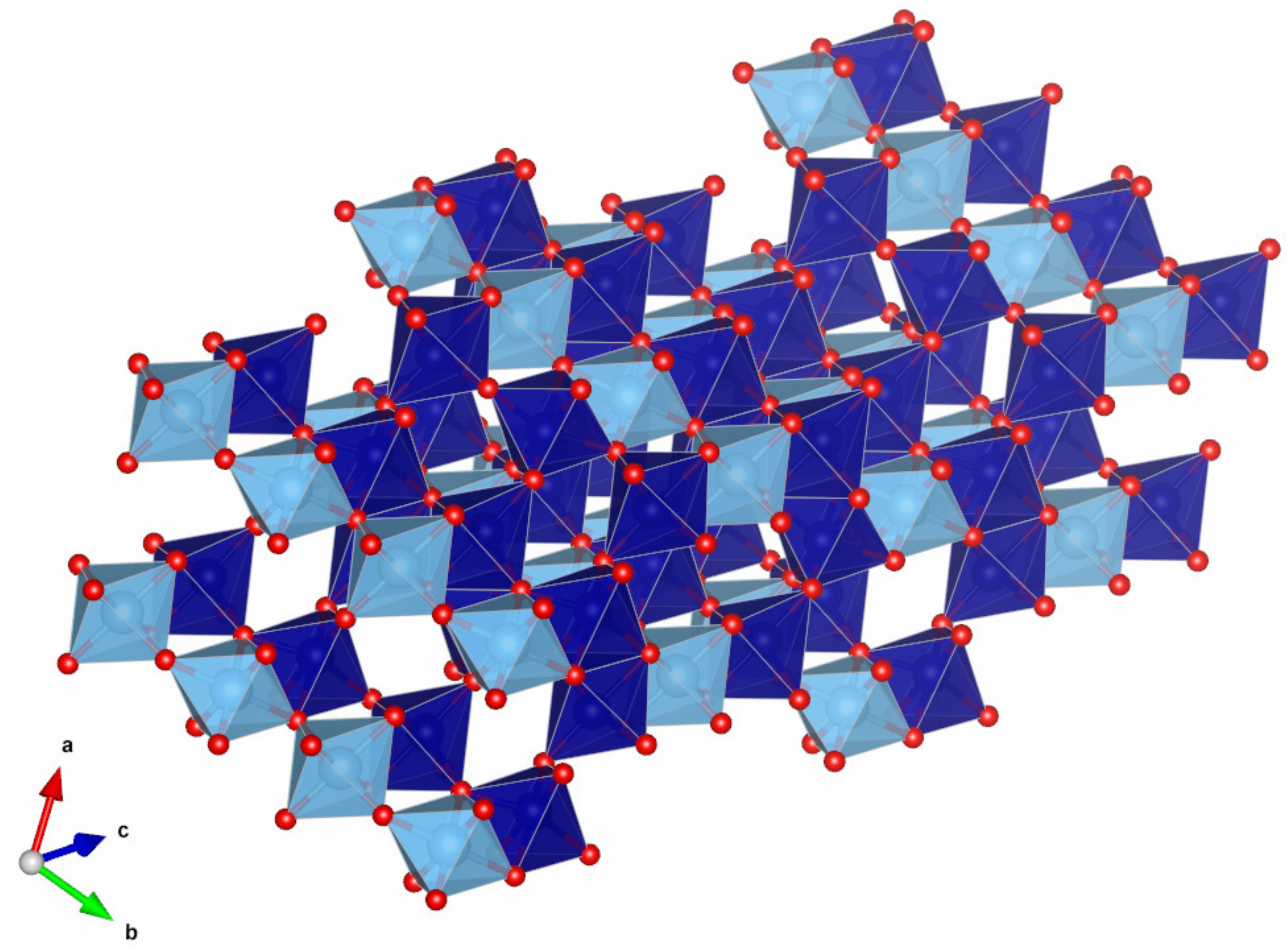}
\caption{(color online) Effective two-Ti-sublattice representation of $\gamma$-Ti$_3$O$_5$
(see text). Light(Dark) blue octahedra correspond to Ti sites on sublattice 1(2).}
\label{fig:charge}
\end{figure}

The oxygen deficiency quite naturally introduces symmetry-inequivalent Ti sites with
potentially different electron occupation. Figure~\ref{fig:ti3o5}b shows four different
Ti sites with especially the Ti4 site surrounded only by five oxygen atoms. The 
$\gamma$-Ti$_3$O$_5$ structure formally has eight Ti sites different by symmetry, which were
also differently treated in our calculations. In the discussion however, to a very good 
approximation, four Ti classes may be grouped to a single effective class, respectively, 
since the local-orbital structure only marginally differs. Figure~\ref{fig:charge} displays
the present Magn{\'e}li structure in that effective two-Ti-sublattice picturing. An obvious
pattern is derived therefrom. The TiO$_6$ octahedra of Ti sites within a given sublattice
are corner sharing, whereas the octahedra are edge sharing between both sublattices. Therefore
the intra-sublattice Ti-Ti distance is about 0.5\,\AA\, longer than the inter-sublattice pair
distance. Both effective sublattices are not of equal-site size, since the Ti2
sublattice (dark-blue octahedra in Fig.~\ref{fig:charge}) covers twice as many Ti sites as the 
Ti1 sublattice (light-blue octahedra in Fig.~\ref{fig:charge}). 
Note that the Ti1 octahedra mark the middle of the stoichiometric undisturbed TiO$_2$ rutile
slabs, which consist of three octahedra in the rutile (001) direction and are interrupted
by a (121) V$_{\rm O}$ defect plane.

Tables~\ref{tab:occ1} and~\ref{tab:occ2} provide the Ti($3d$) occupations for both considered
Ti$_3$O$_5$ structures. Nonsurprisingly in defect rutile, the charge on the Ti4 sites with 
5-fold oxygen coordination is largest, marking that site with a Ti$^{2.8+}$ oxidation state. 
The nearby Ti3 sites are still close to Ti$^{3+}$, while Ti1-2 are closer to Ti$^{4+}$. 
There are no dramatic differences between the numbers based on GGA and those from DFT+DMFT, 
only the occupation of Ti4 is still somewhat higher with correlations. 

The Magn{\'e}li phase $\gamma$-Ti$_3$O$_5$ has a more subtle Ti ordering. The geometrical 
constraints of keeping the TiO$_6$ octahedra in line with the nominal Ti$^{3.33+}$ state leads 
to slightly different charge states on both effective sublattices. The smaller Ti1 sublattice
carries already on the GGA level more charge, with a disproportionation 
$\rho({\rm Ti1})-\rho({\rm Ti2})=0.11$\,e$^{-}$. Electronic correlations increase this charge
splitting only by small amounts, even when invoking a rather large Hubbard $U=8\,$eV. 
Importantly, the system remains always metallic. Also because a seemingly 
straightforward Ti$^{3+}$/Ti$^{4+}$ splitting, as in principle possible in the 
Ti$_4$O$_7$ Magn{\'e}li system with a nominal Ti$^{3.5+}$ oxidation state, is not an obvious
option. Note however that there is also a metal-insulator transition reported for Magn{\'e}li 
Ti$_3$O$_5$,~\cite{tan15} but from $\gamma$-Ti$_3$O$_5$ to $\delta$-Ti$_3$O$_5$ upon lowering 
temperature. Thus further structural changes are in addition indispensable to allow for 
insulating behavior.

\begin{table}[t]
\begin{ruledtabular}
\begin{tabular}{l|cc}
        &  GGA   & DFT+DMFT($U=5\,$eV)  \\ \hline
Ti1     &  0.22  &   0.20         \\ 
Ti2     &  0.26  &   0.24         \\ 
Ti3     &  0.90  &   0.86         \\ 
Ti4     &  1.15  &   1.23         \\ 
\end{tabular}
\end{ruledtabular}
\caption{Comparison of the $3d$ filling of the different Ti sites in defect-rutile
Ti$_3$O$_5$ (see Fig.~\ref{fig:ti3o5}b for Ti labeling). 
The DFT+DMFT data is taken at $T=290\,$K.}\label{tab:occ1}
\begin{ruledtabular}
\begin{tabular}{l|ccc}
        &  GGA   & \multicolumn{2}{c}{DFT+DMFT}     \\ \hline
        &        &   $U=5\,$eV  &   $U=8\,$eV \\
Ti1     &  0.740  &   0.752       &   0.766 \\
Ti2     &  0.630  &   0.624       &   0.617 \\
\end{tabular}
\end{ruledtabular}
\caption{Comparison of the $3d$ filling on the two effective sublattices in
$\gamma$-Ti$_3$O$_5$ (see Fig.~\ref{fig:charge} for Ti-sublattice labeling). 
The DFT+DMFT data is taken at $T=290\,$K.}\label{tab:occ2}
\end{table}

\subsection{Summary and discussion\label{sec:dis}}
We presented a detailed first-principles many-body investigation of the effect of oxygen
vacancies in rutile TiO$_2$, both in the lower- and the higher-concentration regime. In the
former case, the DFT+DMFT results directly provide the three key ingredients known from experiment 
for TiO$_{2-\delta}$, namely semiconducting behavior, shallow levels as well as deep levels. Our
deep-level (or in-gap) positioning is in excellent agreement with results from various experimental 
studies. Thereby the many-body perspective provides a different viewpoint on the longstanding
discussion of deep vs. shallow levels for oxygen-deficient TiO$_2$. Here, both levels are connected
in the many-body sense, similar as a lower Hubbard band and a renormalized quasiparticle state in
a conventional moderately correlated metal. Still the correlation-induced gap opening in the
band-like initial shallow levels is not of standard Mott type. Key to the electron localization
at low V$_{\rm O}$ concentration is the Coulomb-repulsion region between defects, that blocks
the fragile hopping paths. Therefore, the resulting charge gap is rather small and does not
scale with $U$ in a conventional way.

By invoking an excited correlated subspace we brought the semiconducting solution into the 
transport regime, therewith showing that indeed the Coulomb interactions on the Ti sites distant 
from the oxygen vacancies dominantly control the competition between itinerancy and localization.
Moreover, the localized states at the V$_{\rm O}$ display a different occupation and 
energy state upon excitation. This connects our study to other (static-correlation) theory work 
for TiO$_{2-\delta}$, where e.g. different V$_{\rm O}$ charge states and polaron formation are 
discussed.~\cite{mat08,mat10,mor10,jan10,dea12,jan13,lin15,ber15,vas16}

In comparison to oxygen-deficient SrTiO$_{3}$, the defect-rutile problem differs in two crucial
point. First, the defect structure with a V$_{\rm O}$ in nearest-neighbor distance to {\sl three} 
Ti sites, contrary to {\sl two} Ti sites in the perovskite structure, fosters a modified 
oxygen-vacancy impact. The higher connectivity with the defect gives way to the formation of more 
band-like impurity states on the single-electron level. Second, the higher entanglement between
$t_{2g}$ and $e_g$ states lifts the electron dichotomy observed on the SrTiO$_{3-\delta}$
surface,~\cite{lec16} suppressing both, a decoupled $t_{2g}$ band formation and a rather disconnect 
straightforward $e_g$ deep-level formation. Therefore at small V$_{\rm O}$ concentrations, rutile 
TiO$_2$ remains insulating (or semiconducting) due to correlation-induced gap opening in 
($t_{2g}$, $e_g$)-entangled band-like impurity levels. Note in this context however that the 
very correlation details of {\sl bulk} SrTiO$_{3-\delta}$ still need to be investigated by
similar means

In the higher V$_{\rm O}$-concentration regime we focussed on the Ti$_3$O$_5$ stoichiometry,
studying the metallic Magn{\'e}li $\gamma$-phase as well as a theory-derived metallic 
defect-rutile phase. The interplay of Ti coordination and $(t_{2g},e_g)$ electronic structure 
properties were discussed and therefrom the differences in total energy and in the spectral features 
explained. While $\gamma$-Ti$_3$O$_5$ turns out as a 'textbook'-like correlated metal, the 
defect-rutile version displays spectral-weight transfers to higher energies in a much broader 
(incoherent) fashion.
Charge disproportionation is a natural by-product of the V$_{\rm O}$ ordering in rutile-based
TiO$_2$ (which formally includes the  Magn{\'e}li phases)~\cite{lib08} and is already captured 
on the Kohn-Sham level. Electronic correlations beyond the latter provide at least for 
Ti$_3$O$_5$ only minor changes to the Ti charging states. 

Although in direct comparison, defect-rutile Ti$_3$O$_5$ is energetically rather unfavorable 
compared to the Magn{\'e}li phase, its discussion is nonetheless revelant in view of engineering 
memristive processes and devices. Since there in TiO$_2$, transport of V$_{\rm O}$s and electrons 
when starting from the rutile structure is a possible technological aspect. In fact, in realistic 
close-to-device-like TiO$_2$ materials, the formation of defect-rutile {\sl and} Magn{\'e}li 
Ti$_n$O$_{2n-1}$ phases may happen in parallel within a given materials system.~\cite{kwo10} 
Thus understanding the differences between both structural types on a basic level is of 
crucial importance. In the long run, the present study shall contribute to pave the road for 
elucidating further engineering options in the interplay between the transport of oxygen 
vacancies and electrons.

It would also be highly interesting to extent the present investigation to the rutile VO$_2$
system. As the vanadium ion in the $+4$ oxdidation state has $3d^1$ occupation, it is known
that already the stoichiometric system is prone to strong correlation 
physics,~\cite{zyl75,som78,ric94,bie05,naj17}
possibly giving reason to the hallmark metal-to-insulator transition slightly above room 
temperature.~\cite{mor59} Recent doping studies of VO$_2$ via oxygen vacancies display
opportunities to tune the competition between the metallic and insulating 
regime.~\cite{jeo13,zha17}

Finally coming back to basic features of V$_{\rm O}$s in transition-metal oxides, our examination
challenges the simplest views on in-gap states, namely the weak-coupling defect-level
and the strong-coupling Hubbard-band paradigm. Albeit one encounters features of both original
mechanisms in oxygen-deficient TiO$_2$, as in related systems such as SrTiO$_{3-\delta}$ or
the LaAlO$_3$/SrTiO$_3$ interface, a unique and well-defined picture describing the general
nature of defects in transition-metal oxides is still missing. Crystal-field effects, 
renormalizations, Hubbard-band formation, lifetime effects, $p$-$d$ hybridization, screening, 
charge transfer, polaron formation, etc. are potentially part of this demanding physics. 
Thoughts trying to put the problem in an adequate model setting have been put forward from 
several perspectives, e.g. Ref.~\onlinecite{hal76}. The future task is to cast those 
into a sound and solid materials-dependent picture.

\begin{acknowledgments}
We gratefully acknowledge financial upport from the German Science Foundation (DFG) via 
SFB986 and through FOR1346.
Computations were performed at the University of Hamburg and the JURECA 
Cluster of the J\"ulich Supercomputing Centre (JSC) under project number hhh08.
\end{acknowledgments}

\bibliographystyle{apsrev}
\bibliography{bibextra}

\end{document}